\RequirePackage{fix-cm}
\documentclass[twocolumn,epjc3]{svjour3}  
\usepackage{xcolor}%
\usepackage{amsmath}%
\usepackage{multirow}
\usepackage{multicol}
\smartqed  % flush right qed marks, e.g. at end of proof
\RequirePackage{graphicx}
\newcommand{\VNTL}{HV}
\newcommand{\Thickness}{200~$\mu$m}
\begin{document}

\title{NTL-amplified cryogenic light detectors with optically transparent electrodes}

%\titlerunning{Short form of title}        % if too long for running head

\author{Matteo Biassoni\thanksref{e1,addr1}\and
        Andrea Nava\thanksref{e2,addr2,addr1}\and
        Oscar Azzolini\thanksref{addr3}\and
        Mattia Beretta\thanksref{addr1}\and
        Tommaso Bradanini\thanksref{addr2,addr1}\and
        Chiara Brofferio\thanksref{addr2,addr1}\and
        Paolo Carniti\thanksref{addr2,addr1}\and
        Simone Copello\thanksref{addr4}\and
        Mourad El Idrissi\thanksref{addr3}\and
        Marco Faverzani\thanksref{addr2,addr1}\and
        Elena Ferri\thanksref{addr1}\and
        Massimo Girola\thanksref{addr2,addr1}\and
        Luca Gironi\thanksref{addr2,addr1}\and
        Claudio Gotti\thanksref{addr1}\and
        Léonard Imbert\thanksref{addr1}\and
        Giorgio Keppel\thanksref{addr3}\and
        Nicola Manenti\thanksref{addr5,addr4}\and
        Ilaria Molinari\thanksref{addr2,addr1}
        Irene Nutini\thanksref{addr1}\and
        Maura Pavan\thanksref{addr2,addr1}\and
        Daniele Peracchi\thanksref{addr2,addr1}\and
        Gianluigi Pessina\thanksref{addr1}\and
        Sonja Schneidewind\thanksref{addr1}\and
        Davide Trotta\thanksref{addr2,addr1}
}

%\thankstext{t1}{Grants or other notes
%about the article that should go on the front page should be
%placed here. General acknowledgments should be placed at the end of the article.
\thankstext{e1}{e-mail: matteo.biassoni@mib.infn.it}
\thankstext{e2}{e-mail: andrea.nava@mib.infn.it}

%\authorrunning{Short form of author list} % if too long for running head

\institute{
    INFN, Sezione di Milano Bicocca, Milano, 20126, Italy\label{addr1}
    \and
    Dipartimento di Fisica "G.Occhialini", Università degli Studi di Milano - Bicocca, Milano, 20126, Italy\label{addr2}
    \and
    INFN, Laboratori Nazionali di Legnaro, Legnaro (PD), 35020, Italy\label{addr3}
    \and
    INFN, Sezione di Pavia, Pavia, 27100, Italy\label{addr4}
    \and
    Dipartimento di Fisica A.Volta, Università di Pavia, Pavia, 27100, Italy\label{addr5}    
}

\date{Received: date / Accepted: date}
% The correct dates will be entered by the editor

\maketitle

\begin{abstract}
Neganov-Trofimov-Luke (NTL) effect is used by experiments based on cryogenic detectors to boost the sensitivity of light-sensitive devices down to a few optical photons.
In this work we introduce a silicon light‑detector technology that implements NTL amplification at millikelvin temperatures using transparent indium–tin–oxide (ITO) electrodes.
The ITO electrodes enable an electric field perpendicular to the wafer surface, mitigating surface charge recombination, and thanks to their optical properties simultaneously serve as an anti‑reflective coating.
By combining these two functions in a single element, the fabrication process is simplified, yielding more robust and cost‑effective devices.
We report on the production and characterization of the first batch of these detectors.
We performed a room-temperature characterization of the ITO electrodes, verifying the structural and optical characteristics of the deposited electrodes.
We then operated 2 of these devices as cryogenic calorimeters at millikelvin temperatures.
Finally, we develop a consistent analytical model for the NTL gain for both ionizing particles and optical photons, successfully describing the gain dependence on the NTL bias and explicitly accounting for the partial electrode coverage of the device surface.
\keywords{Indium tin oxide \and low-temperature detector \and Neganov-Trofimov-Luke effect \and neutrino \and double-beta decay \and rare events \and light detector}
% \PACS{PACS code1 \and PACS code2 \and more}
% \subclass{MSC code1 \and MSC code2 \and more}
\end{abstract}

\section{Introduction}\label{introduction}
Cryogenic calorimeters operated at milliKelvin temperature are a key component of many modern particle and astro-particle physics experiments \cite{Giuliani:2012a}. In particular, experiments searching for neutrino-less double beta decay with cryogenic calorimeters \cite{biassoni:2020} couple these devices to a main calorimeter composed of a crystal embedding the isotope candidate to the decay. 
By detecting the light produced by an interaction in the main crystal, whether via scintillation or Cherenkov emission, these detectors can distinguish between particle types. In particular, background-inducing alpha events can be effectively discriminated, preserving only signal-like beta interactions.
Such a capability is also of interest for the search of direct Dark Matter (WIMPs) interactions through nuclear recoil detection with cryogenic calorimeters \cite{Abdelhameed:2019hmk}.
The basic element of these cryogenic light detectors is a thin O(100 $\mu$m) wafer acting as photon absorber, typically made of high purity silicon, germanium or sapphire. This absorber is equipped with a temperature transducer, converting the increase in temperature generated by absorbed photons into an electric signal. Many options exists for the temperature transducer, depending on the specific application.
Devices like Transition Edge Sensors \cite{Schaffner2015,Singh:2023}, Kinetic Inductance Detectors \cite{Battistelli:2015vha} or Metallic Magnetic Calorimeters \cite{Gray2016} provide high sensitivity and fast response times, exploiting promising but complex technologies. In particular, the complex readout system for these devices pose significant technological challenges to the increase of number of channels operated simultaneously.  
On the other hand, Neutron Transmutation Doped Germanium thermistors (NTD) provide a less sensitive but more robust solution. NTDs are coupled to the photon absorber with a thin layer of adhesive \cite{Alfonso:2023a} and can be read-out with conventional front-end electronics, being high resistivity devices. In addition, their deployment on the scale of thousands of channels in low-radioactivity setups like CUORE \cite{Adams:2025} has been demonstrated.
Cryogenic light detectors using Ge as absorber and NTDs as temperature transducers have already been successfully deployed in the CUPID-0 \cite{Azzolini:2018tum} and CUPID-Mo \cite{Armengaud:2020} experiments. Both these detectors demonstrated the possibility to perform alpha-beta discrimination to reduce the background in the search of neutrino-less double beta decay \cite{Azzolini:2022,Augier:2022}.
The relatively limited sensitivity of NTDs, however, makes them suboptimal for applications where few photons need to be detected, as they restrict the ultimate performance required by next-generation experiments like CUPID \cite{Cupid:2025avs}.
To overcome this limitation, it is possible to exploit the Neganov-Trofimov-Luke (NTL) effect in cryogenic calorimeters \cite{Luke:1988}. By applying a static electric field to the photon absorber, the electron-hole pairs generated upon photons absorption are drifted in the detector volume. As long as this drift happens without charge recombination, the additional kinetic energy provided to the system gets converted into heat, amplifying the thermal signal only. Within these assumptions, this additional thermal energy is proportional to the voltage and independent from the electric field magnitude. In real implementations, the electric field magnitude and geometry can affect the probability of charge recombination and thus the achievable gain. 
State-of-the-art light detectors implementing the NTL effect have been used by the CUPID experiment from its initial R\&D phase to its technology demonstrators \cite{Armatol:2025zrq,Gironi:2016nae,Biassoni:2015eij,Willers2015a,Pattavina2016,Novati:2019,Chernyak2016,Berge:2017nys}. In these detectors the electric field is generated by applying a DC voltage to a pattern of concentrical electrodes deposited on one of the surfaces of a germanium wafer. The electrodes are connected, in an alternating scheme, by aluminum bonding wires, so that only two electrodes need to be connected to the externally provided bias voltage. These detectors need to operate at a bias voltage of 100 V or larger to provide the required amplification factor \cite{Armatol:2025zrq}.
To meet this requirement, the surface quality of the wafer must be carefully monitored. Given the geometry of these electrodes, the electric field is mainly parallel to the wafer surface, as is the drift of the charges. This makes these devices extremely sensitive to surface quality. Even the slightest contamination before or during the electrodes manufacturing can in fact lead to the development of leakage currents at voltage smaller than the target.
Moreover, the probability of recombination with surface defects is non-negligible, as the electron-hole pairs drift close to the surface. This leads to thermal signals amplification smaller than the one theoretically achievable. Finally, in order to maximize the absorption of the optical photons in the germanium, a silicon oxide anti-reflective coating must be applied after the deposition of the electrodes, in a separate step that increases the complexity and cost of the production.

In this work we propose a new idea for the construction of NTL-amplified light detectors, in an attempt to overcome the limitations of the currently available technology.

\section{Proposed device: NTL with transparent electrodes}

Our goal is to realize a cryogenic light detector where a strong electrical field can be applied to the bulk of the photon absorber, avoiding surface effects. 
This could be obtained by equipping the wafer used as absorber with electrodes as wide as the surface, creating a planar capacitor with the wafer bulk as the internal dielectric. 
Within this idea, the application of metallic electrodes is not feasible, since it will prevent optical photons to be absorbed and transformed into a detectable thermal signal. 
A solution to this problem comes from the transparent conductive films, developed and engineered in the definition of more efficient solar cells \cite{nano13071226}.
These materials are used to create electrodes capable of collecting the photovoltaic currents without interfering with photon absorption. While different materials exist for this application  \cite{Ding2024_Metal_Nanowire_TE_Review,Wang2021_Flexible_TE_Review,Li2020_AgNW_Review_JMCC}, one of the mostly used and studied is Indium Tin Oxide (ITO) \cite{Patel2024_TCF_Review,ITO_Rajendran}.
This material, typically deposited by sputtering, allows to create electrodes with good conductive properties (Bulk resistivity $\sim2\cdot10^{-4} \Omega\cdot\text{cm}$ \cite{nano13071226}) and optical characteristics tunable by modifying the thickness of the oxide layer ($\sim$97\% transmittance at 550~nm  \cite{ITO_Rajendran}).

The design we propose revolves around state-of-the-art cryogenic light detectors with wide ITO electrodes deposited by sputtering on opposing faces.
This technology can ensure high bulk electrical field, providing efficient NTL amplification.
By optimizing the thickness of the ITO deposition, the electrode can also double as anti-reflective coating, providing efficient photon detection.

The practical implementation that we present in this work revolves around instrumenting a high purity silicon wafer with ITO transparent detectors deposited by sputtering. 
The use of silicon instead of germanium for this first test is motivated by the lower cost of the raw material, together with the excellent results reported with silicon detectors with a more classical electrode design  \cite{Biassoni:2015eij}.
In particular, the wafers are high purity silicon with diameter of 2-inches (100) orientation, resistivity $>$10 k$\Omega\cdot$cm  and thickness \Thickness{}. 
The ITO electrodes are circular with a diameter of 40 mm, so that a ring up to 5.4 mm from the edge of the wafer remains uncovered. This is done to minimize the probability of a parasitic current developing around the edge of the wafer. In the future, further optimization could lead to a reduction of the size of this area, improving the effective surface of the detector. A golden pad is also deposited on each detector surface. These pads partially overlap with the ITO electrode and are meant to host the wire bonding through which the NTL voltage is applied.

The production will be described in Section \ref{production} and will be followed by the optical (Section \ref{optical characterization}) and low temperature (Section \ref{crio-characterization}) characterization. 

\begin{figure}[h]
    \centering
    \includegraphics[width=0.47\textwidth]{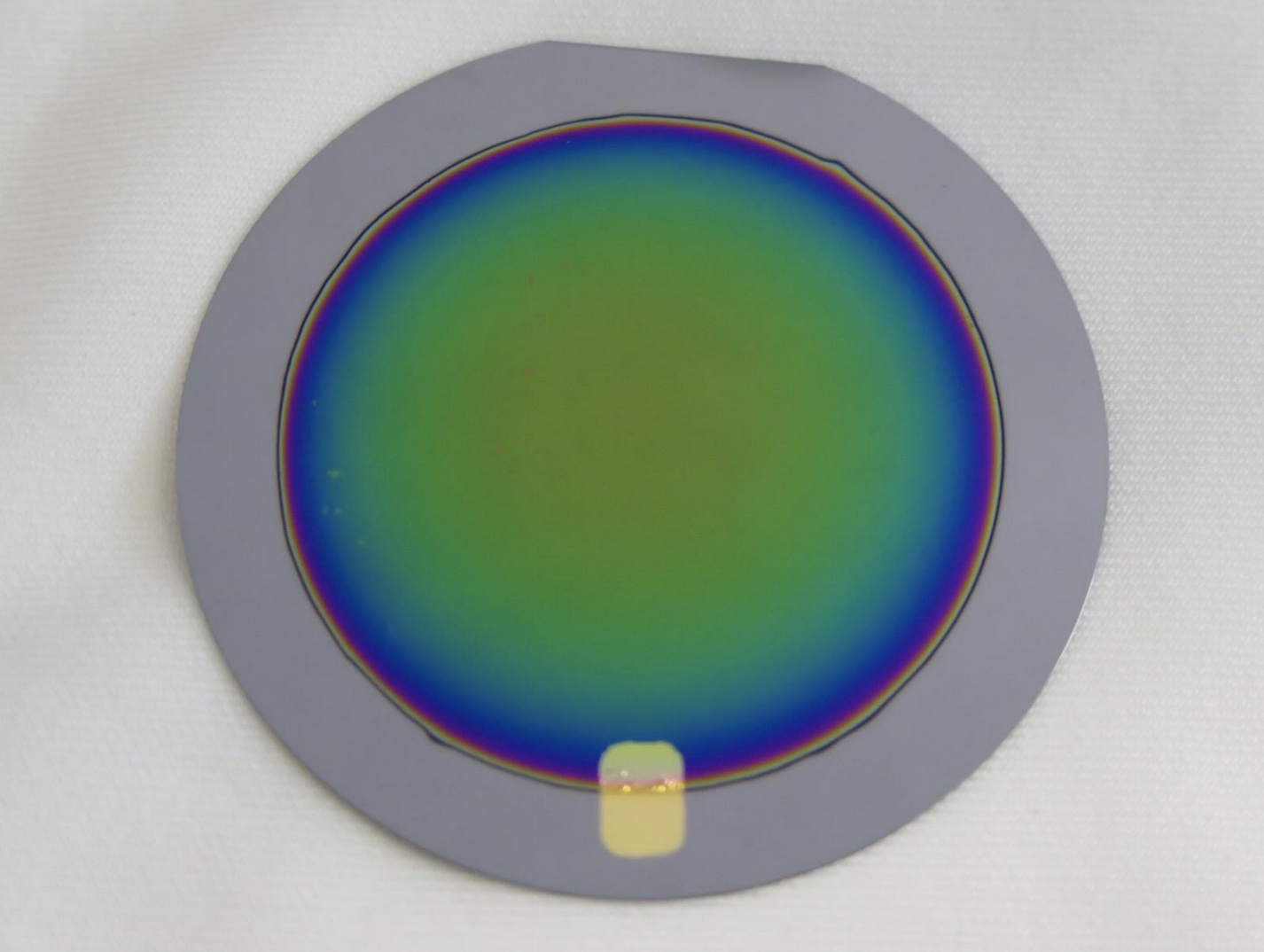}
    \includegraphics[width=0.47\textwidth]{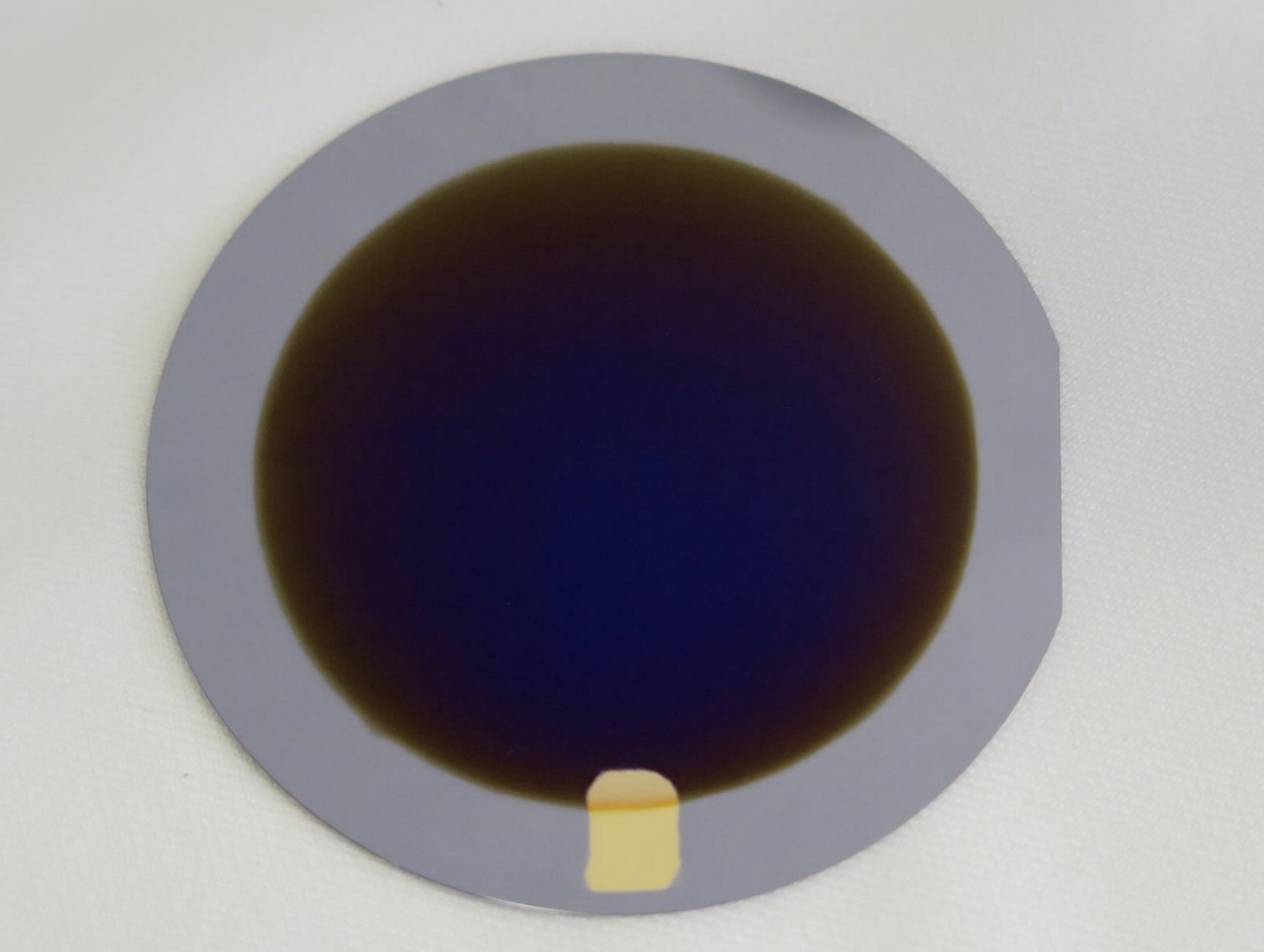}
    \caption{Pictures of the two devices tested in this work (ITO1 on the top, ITO4 on the bottom). The ITO electrodes are clearly visible, as well as the golden bonding pads. The different color of the electrodes is due to their different thickness, affecting their optical properties.}\label{fig:devices}
\end{figure}

The produced devices that are tested in this work will be referred to as ITO1 and ITO4 and shown in Figure~\ref{fig:devices}.

\subsection{ITO electrodes production}\label{production}
The process to produce ITO electrodes is based on sputtering deposition \cite {TSENG2013157} and was carried out in a cylindrical stainless-steel high-vacuum chamber with an internal diameter of 34~cm and a length of 40~cm. The system was evacuated to a base pressure below $5 \times 10^{-6}$~mbar without venting between consecutive depositions, using a Pfeiffer turbomolecular pump (nominal pumping speed: 75~l\,s$^{-1}$) backed by an Edwards TriScroll dry scroll pump (nominal pumping speed: 10~m$^{3}$\,h$^{-1}$).
The relatively low base pressure ensured minimization of residual gas contamination, particularly water vapor, which are known to significantly influence stoichiometry and electrical transport properties in ITO thin films.
The substrate holder was electrically grounded and made of stainless steel.
Substrates were mounted on a planar holder positioned 9~cm below the cathode in a top-down configuration, with the magnetron source located at the top of the chamber and the substrates placed at the bottom. No substrate rotation was employed during deposition.
%The deposition of ITO and gold pads is followed by a post deposition annealing, to improve the film characteristics. 
All deposition parameters are summarized in Table \ref{tab:production parameters}. The description of the deposition procedure is described in the following sections.

\begin{table*}
% table caption is above the table
\caption{Deposition parameters for the ITO electrode and the gold bonding pad.}\label{tab:production parameters}%
% For LaTeX tables use
\begin{tabular*}{\textwidth}{@{\extracolsep{\fill}}lll@{}}
\hline\noalign{\smallskip}
Parameter & ITO electrodes & Au bonding pad  \\
\noalign{\smallskip}\hline\noalign{\smallskip}
Ar flux [sccm] & 16  & 17  \\
Ar Working pressure [mbar] & $8.8 \times 10^{-3}$~  & $9.6 \times 10^{-3}$ \\
Power [W] & 40 & 100 \\
Tension [W] & 400 & 325\\
Current [A] & 0.1 & 0.3 \\
Deposition Rate [nm/min] & 30 & 28 \\
Working Distance [cm] & 10 & 9 \\
Base pressure [mbar] & $<5.0 \times 10^{-6}$~ & $<5.0 \times 10^{-6}$ \\
\noalign{\smallskip}\hline
\end{tabular*}
\end{table*}

\subsubsection*{ITO Deposition Procedure}

ITO films were deposited by DC magnetron sputtering using a 4-inch balanced planar magnetron cathode equipped with a commercial 4N purity ITO target (90~wt.\% In$_2$O$_3$ -- 10~wt.\% SnO$_2$).
The working pressure during deposition was maintained at $9.6 \times 10^{-3}$~mbar with high-purity argon (99.999\%) supplied as sputtering gas at a constant flow rate of 17~sccm; no additional oxygen was introduced into the plasma.
All depositions were carried out at room temperature without intentional substrate heating.
Under these conditions, we measured an approximate deposition rate of 28~nm\,min$^{-1}$. This value is determined measuring ex-situ the thickness of a test deposition with a contact profilometer, and diving this quantity by the deposition time.

\subsubsection*{Gold pad deposition}

Gold contact pads were deposited for device fabrication using a similar magnetron sputtering system equipped with a 2-inch gold target. The deposition was performed at a power of 40 W under an argon working pressure of $8.8 \times 10^{-3}$~mbar.
A gold thickness of 300 nm was deposited without the use of any adhesion layer (e.g., Ti or Cr). The absence of an intermediate layer was chosen to avoid potential interface diffusion or additional stress contributions that could affect cryogenic performance.

\subsection{Optical characterization}\label{optical characterization}

For this production, the ITO thickness was targeted to absorb photons at the peak value of lithium molybdate emission (550-600 nm), the scintillator chosen for the CUPID experiment \cite{Cupid:2025avs}. 
To verify that the target was met, we performed reflectance measurements of the ITO coatings deposited on both sides of the silicon wafer.
These measurements were performed at room temperature, probing both the center of and the edge of the coating with a spectrophotometer (Konica Minolta and CM-2600d). By performing the measurements at two different positions, deposition non-uniformity associated with the magnetron sputtering source can be evaluated. From the visual inspection of the samples shown in Figure~\ref{fig:devices}, a difference is expected, since the samples show a radial color change.

\begin{figure}[h]
    \centering
    \includegraphics[width=0.5\textwidth]{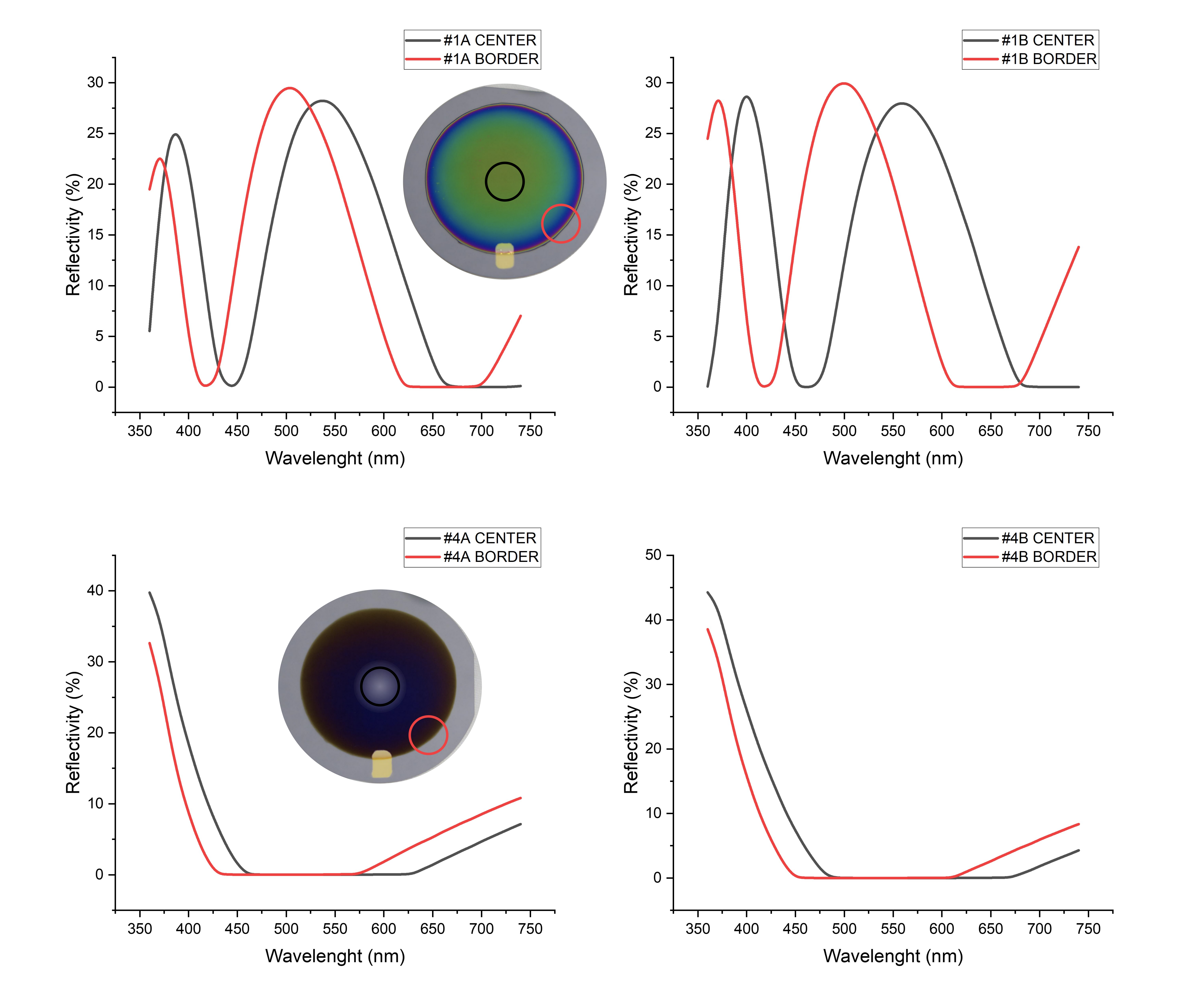}    
    \caption{Reflectivity curves of the ITO1 (top) and ITO4 (bottom) samples tested on both sides of the silicon wafer (A and B, left-right) and in two different position of the coated area: center (black) and border (red).
    The electrodes were designed to avoid the reflection of light in the 500-600~nm range, corresponding to Lithium Molybdate emission. This objective is demonstrated for ITO4 and only partially for ITO1.
    The comparison between sides and positions demonstrates some variation, as expected from the visible diffraction patterns.}\label{fig:reflectivity}
\end{figure}

The results of these measurements are reported in Figure~\ref{fig:reflectivity} for both ITO1 and ITO4 samples.

ITO4 shows the expected reflectivity spectrum, with a minimum in the 500-600~nm region. On the other hand, ITO1 presents a reflectivity spectrum with two peaks typical of thick samples.

The depositions on the two sides of both silicon wafers have similar responses, as well as the corner and central regions. ITO1 shows higher variations, as expected from its appearance.

From these reflectivity spectra, we estimate the ITO coating thickness $d$. 
As the two samples exhibit different behavior, two different formulas had to be used.
ITO1 presents the behavior of a thick samples with two maxima in the reflectivity spectrum, so the following expression was used:
\begin{equation}
d = \frac{\lambda_1\cdot\lambda_2}{2\cdot n_{ITO}\cdot(\lambda_2-\lambda_1)}
\label{eq:d_thicksamples}
\end{equation}
where $d$ is the film thickness, $\lambda_1$ and $\lambda_2$ are the wavelengths of two  maxima and $n_{ITO}$ is the refractive index of ITO. For this wavelength range $1.8<n_{ITO}<2.1$, therefore $d$ was calculated with the average value $n_{ITO} = 1.95$ and an uncertainty of 0.15 was associated \cite {Moerland:16}.

For ITO4, which behaves more like an ideal thin wafer, the following expression was used instead:
\begin{equation}
d = \frac{\lambda_0}{4\cdot n_{ITO}}
\end{equation}

where all parameters are the same as in \ref{eq:d_thicksamples} except $\lambda_0$, the central wavelength of the interval at zero reflectivity. 

The results for this calculation are reported in table~\ref{tab:ITO_PRODUCED}, showing the significantly different values for the two samples. 

\begin{table}
% table caption is above the table
\caption{Summary of the thickness evaluation for the produced devices. The values are calculated from the pattern in the reflectivity spectra - refer to the text for details. The uncertainties have to be read as ranges for the value, dictated by the possible range of ITO reflective indexes (1.8-2.1 \cite{Moerland:16}). On the "Average" column, the range includes also the variability among all different evaluations (two sides and two positions). The thicknesses are significantly different between the two samples.}\label{tab:ITO_PRODUCED}%
% For LaTeX tables use
\centering
\begin{tabular}{ccccc}%{0.5\textwidth}{@{\extracolsep{\fill}}cccc@{}}
\hline\noalign{\smallskip}
\multirow{2}{*}{Detector}   &\multirow{2}{*}{Side} &\multicolumn{3}{c}{Thickness [nm]}\\
                            &                      &   Center  & Edge   & Average\\
\noalign{\smallskip}\hline\noalign{\smallskip}
\multirow{2}{*}{ITO1}   &   A   &   353$\pm$55  & 363$\pm$56    &  \multirow{2}{*}{375$\pm$111}\\
                        &   B   &   376$\pm$58  & 406$\pm$73    &                              \\
                        \hline
\multirow{2}{*}{ITO4}   &   A   &   72$\pm$11   & 63$\pm$10     &  \multirow{2}{*}{69$\pm$21}\\
                        &   B   &   74$\pm$11   & 69$\pm$11     &                            \\
\noalign{\smallskip}\hline
\end{tabular}
\end{table}

\section{Low-temperature characterization}\label{crio-characterization}
Following room temperature optical and structural characterization, the detectors have been operated at milli-Kelvin temperature as cryogenic light detectors.
The goal of the characterization was to demonstrate that the devices are sensitive to optical photons and that thermal gain can indeed be increased via NTL effect by applying an electric field through the silicon bulk.

\subsection{Experimental setup}\label{crio-setup}
\begin{figure}[h]
\centering
\includegraphics[width=0.45\textwidth]{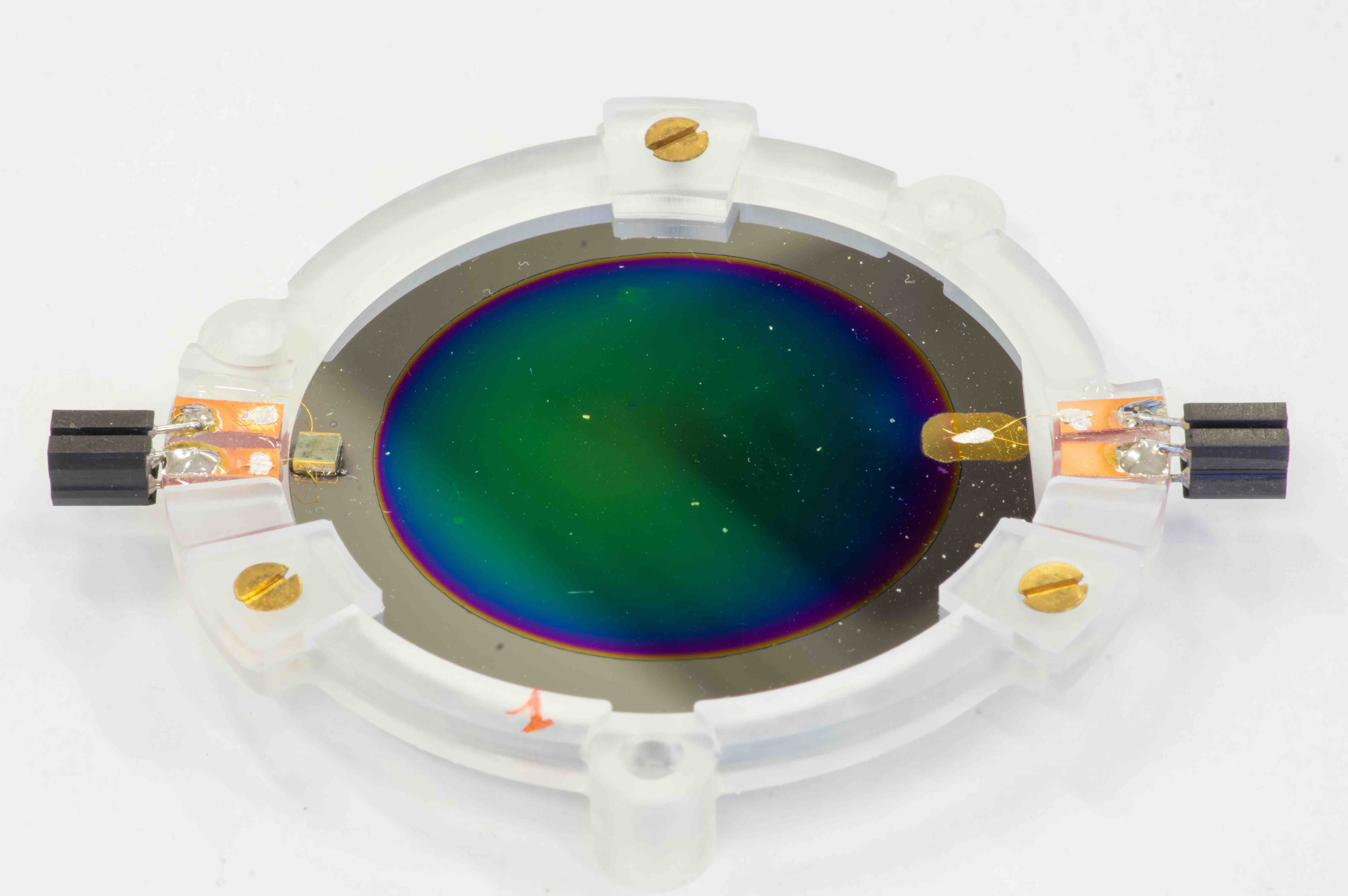}
\includegraphics[width=0.45\textwidth]{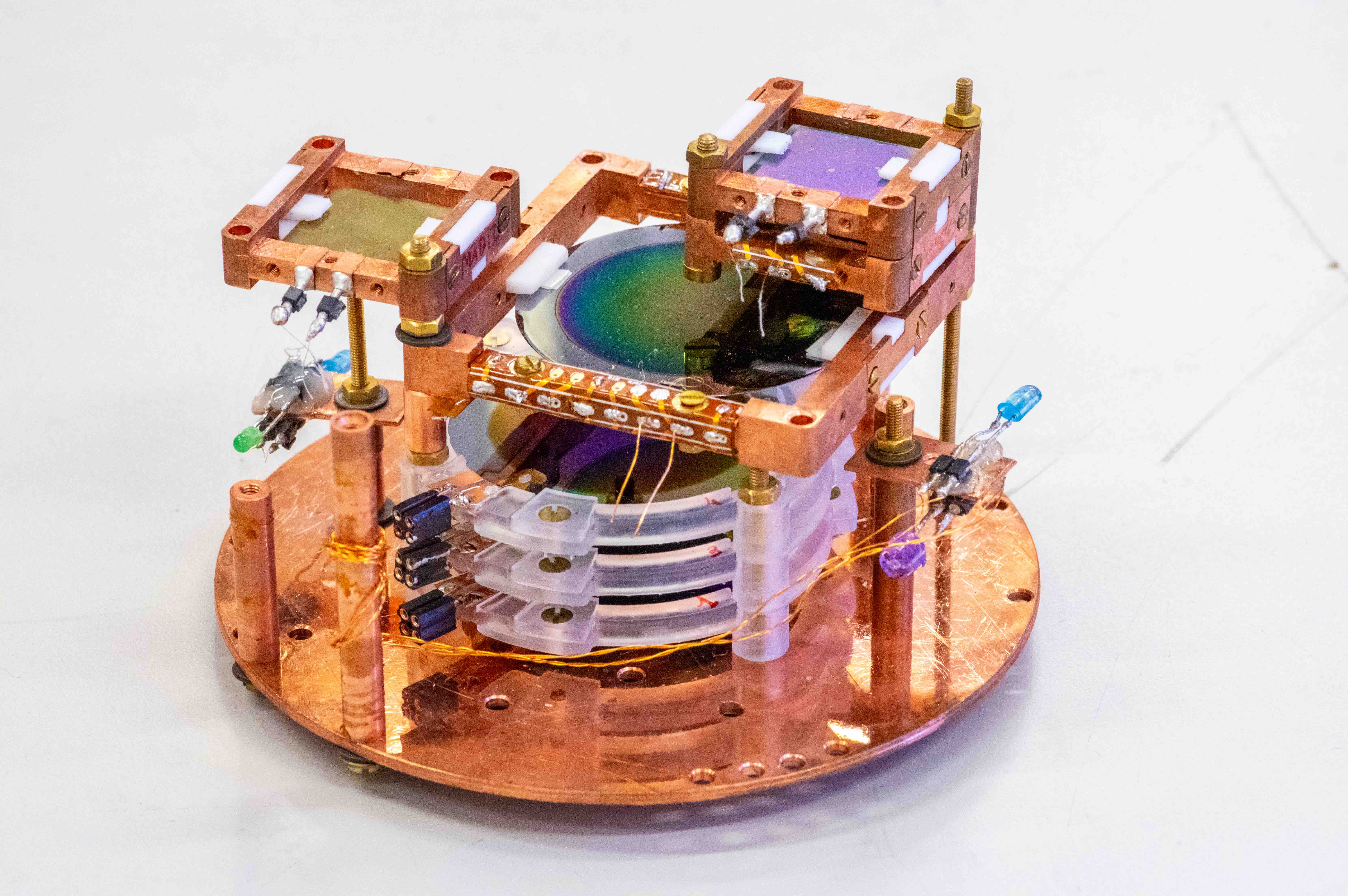}
\caption{Top: ITO1, one of the studied detectors assembled in a plastic frame. Readout and NTL bias gold wires connected to Cu-PEN pads are visible. Bottom: the stack of detectors ready to be attached to the cryostat. Top (ITO4) and bottom (ITO1) detectors in the stack of plastic frames are the ones characterized in this work.}\label{fig:detector}
\end{figure}

In order to operate it as a cryogenic calorimeter, each device has been equipped with a NTD thermistor. The thermistor is glued to the silicon substrate with spots of epoxy glue. We took care of gluing the NTD on the bare silicon, outside the area where the ITO deposition is present.
The detector was assembled in a frame (both plastic and copper frames have been used in subsequent measurements) that provided mechanical support to the wafer, held in place by plastic clamps, as shown in Figure \ref{fig:detector}.
The frames are equipped with flexible PEN or Kapton PCBs that provide an interface between the cryostat wiring and the device electrodes. Specifically, 25~$\mu$m gold wires have been used to connect the NTD electrodes (via ball bonding) as well as the ITO electrodes on both sides of the device (via indium cold bonding). We took care of connecting to ground the ITO electrode on the side where the NTD is glued. NTL bias voltage (either positive or negative) is always applied on the opposite face, minimizing the likelihood of parasitic currents developing through the NTD electrodes, virtually grounded by the front end electronics.

The individual detectors are first assembled in the mechanical frames and then stacked together on top of a copper plate, as shown at the bottom of Figure~\ref{fig:detector}. ITO detectors are assembled with other detectors tested in the same run but outside the scope of this work. The copper plate is the base of the experimental volume, later attached to an Oxford TL-200 dilution refrigerator mixing chamber.
This cylindrical volume, defined by two copper plates and a copper sheet, is equipped with a diffusive optical fiber coupled at the mixing chamber level to a transport fiber from room temperature. With this setup, photon bursts of different wavelength can be generated at room temperature with laser diodes or LEDs, and injected into the experimental volume. By generating light pulses shorter than the detectors time response we can simulate the emission of scintillation or Cherenkov light and evaluate the detector response to these inputs.

Each NTD is connected to a bias and readout channel, based on the CUORE front-end electronics concept  \cite{Arnaboldi:2015a:powersupply,Arnaboldi:2018:frontend}.
For each thermistor there is a bias circuit designed to generate a constant current, independent on the NTD resistance.
This current, usually lower than a few nA, heats the NTD and allows to choose its working resistance, a parameter tied to its sensitivity and response  \cite{Adams:2022b}.
Once the resistance of the NTD is set, a change in the temperature of the absorber is detected as a change in this resistance, measured as a voltage drop across the NTD itself.
This quantity is amplified by a high gain, low noise amplifier \cite{Arnaboldi:2006mx} and digitized by a custom 24 bit, 1 kHz ADC.
The voltage across the NTD, referred to as baseline, is continuously digitized and stored for offline event reconstruction.
A standard analysis pipeline developed for cryogenic calorimeters is used to detect pulses over this baseline, corresponding to instantaneous energy depositions.
These physical events are then characterized through the same pipeline extracting different parameters of interest, such as the amplitude and the rise and decay times \cite{CUORE:2025vno}.

\subsection{Data-taking and analysis}\label{analysis}
\begin{figure}[h]
    \centering
    \includegraphics[width=0.48\textwidth]{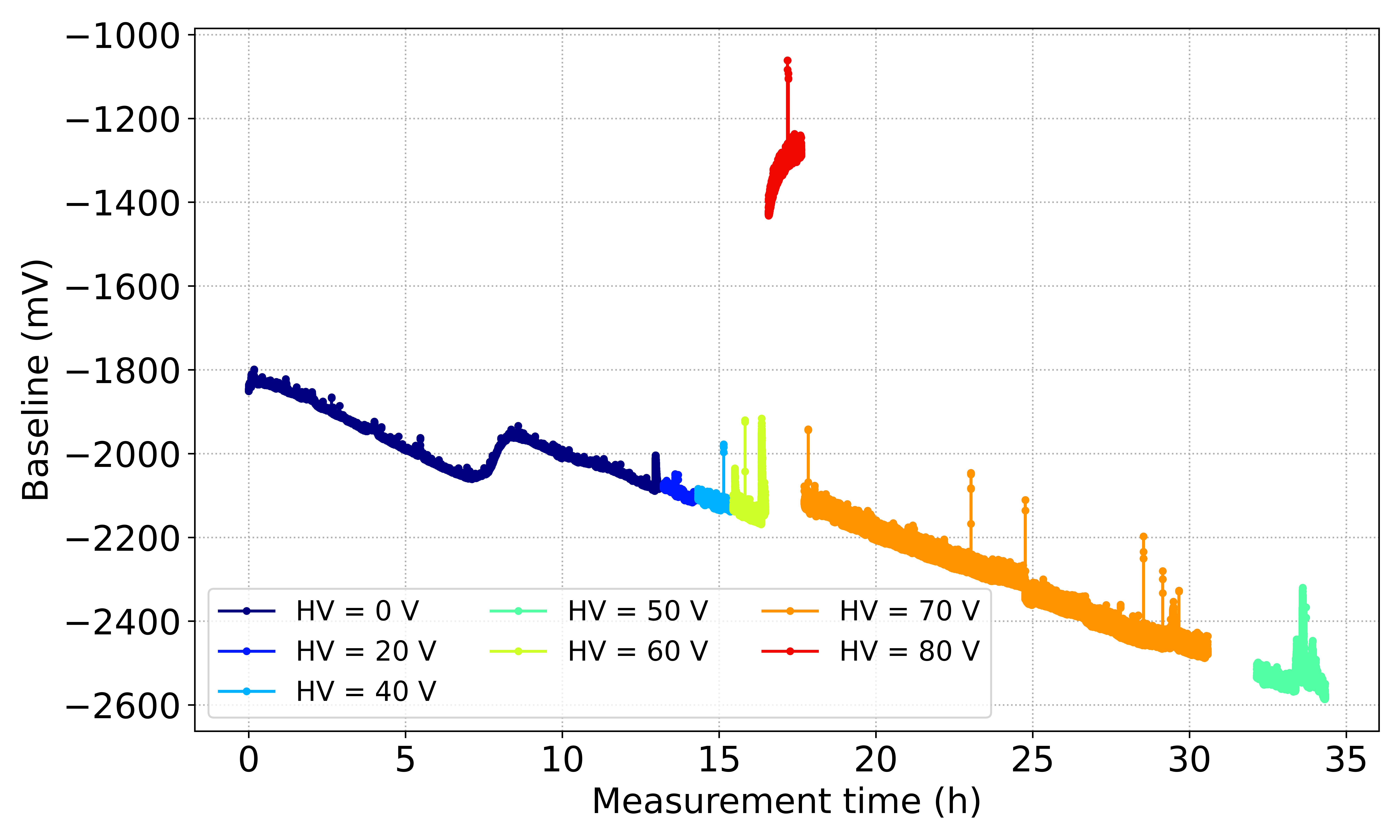}
    \caption{ITO4 baseline as a function of time for different NTL bias voltages (\VNTL). For cryogenic calorimeters, this variable is directly proportional to the temperature of the absorber. A rapid increase when setting \VNTL~=~80 V can be observed, corresponding to the onset of a leakage current heating the detector.}\label{fig:baselineTime}
\end{figure}

The performance of the detectors, in terms of noise level, pulse shape and amplification, have been investigated by applying different NTL voltages (\VNTL) to ITO electrodes.
The first step in the characterization was the evaluation of the highest applicable voltage. This quantity, referred to as breakdown voltage, was evaluated progressively increasing the voltage and monitoring the detector baseline, proportional to the base temperature of the absorber. In case a leakage current develops, the calorimeter gets heated by joule dissipation, causing a baseline increase. 
Such an increase has been observed for \VNTL$>$50 V (\VNTL$>$70 V) for ITO1 (ITO4). 
The plot in Figure \ref{fig:baselineTime} shows the ITO4 baseline value for different sub-runs, for which different bias voltages were applied.
In addition to a general decrease in the baseline value due to the whole system cooling down over time, an off-trend increase is evident when applying \VNTL~=~80~V. 
% It has to be noted the the developed current was small enough to maintain the detector operational and to keep the cryogenic system unaffected.  

\begin{figure}[h]
    \centering
    \includegraphics[width=0.48\textwidth]{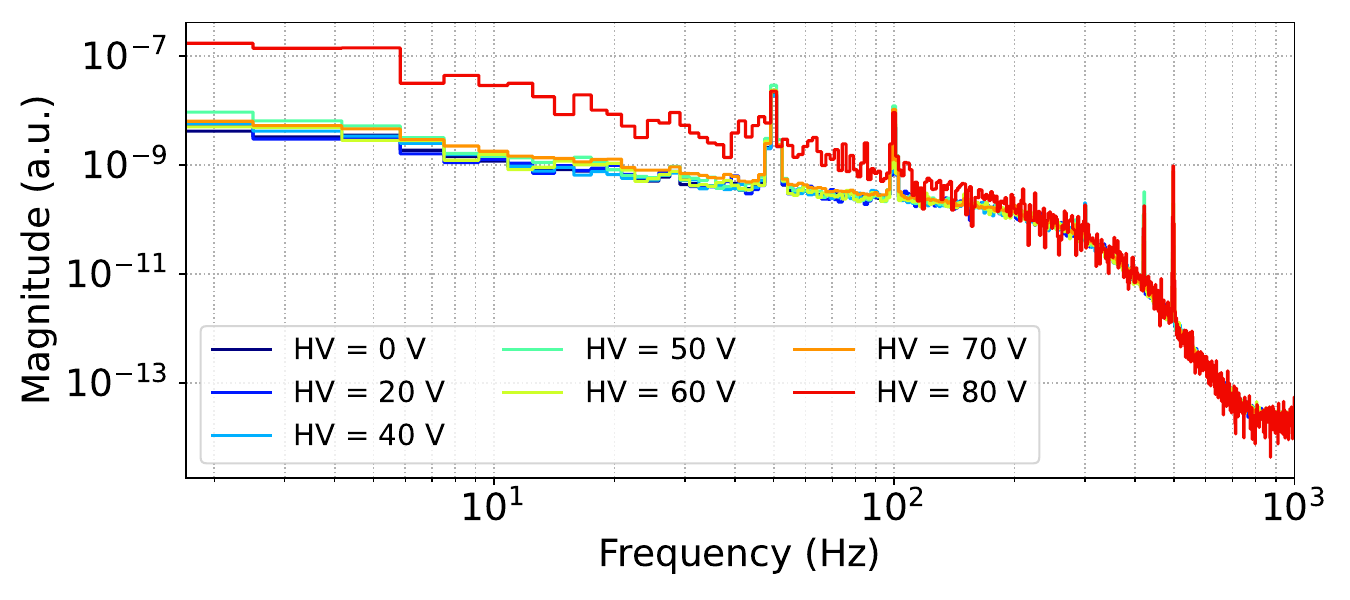}
    \caption{ITO4 Average Noise Power Spectra for different NTL voltages (\VNTL) applied to the devices. The increase of noise for \VNTL~=~80 V (red curve) can be explained as an additional shot noise (low frequency) caused by the movement of charges due to the onset of a leakage current.}\label{fig:HV_effect}
\end{figure}

For the different \VNTL, we evaluated also the Average Noise Power Spectrum (ANPS) for the ITO detectors, selecting portions of the baseline without signals and averaging the modulus of their Fourier transform. As an example, Figure \ref{fig:HV_effect} shows the ANPS for ITO4 for different levels of NTL voltages. 
The noise level is stable for \VNTL{} smaller than the breakdown voltage, while a higher ANPS was measured for \VNTL~=~80~V. The fact that the noise level is the same with and without \VNTL{} means that any observed amplification of the signal due to the NTL effect can be directly translated into an improvement in the signal-to-noise ratio.
The increase of noise above the breakdown voltage is explained as the baseline increase by the onset of the leakage current. The movement of charges inside the volume of the absorber, in fact, produces a shot-like noise, increasing the low frequency components in the ANPS.

\begin{figure}[h]
    \centering
    \includegraphics[width=0.48\textwidth]{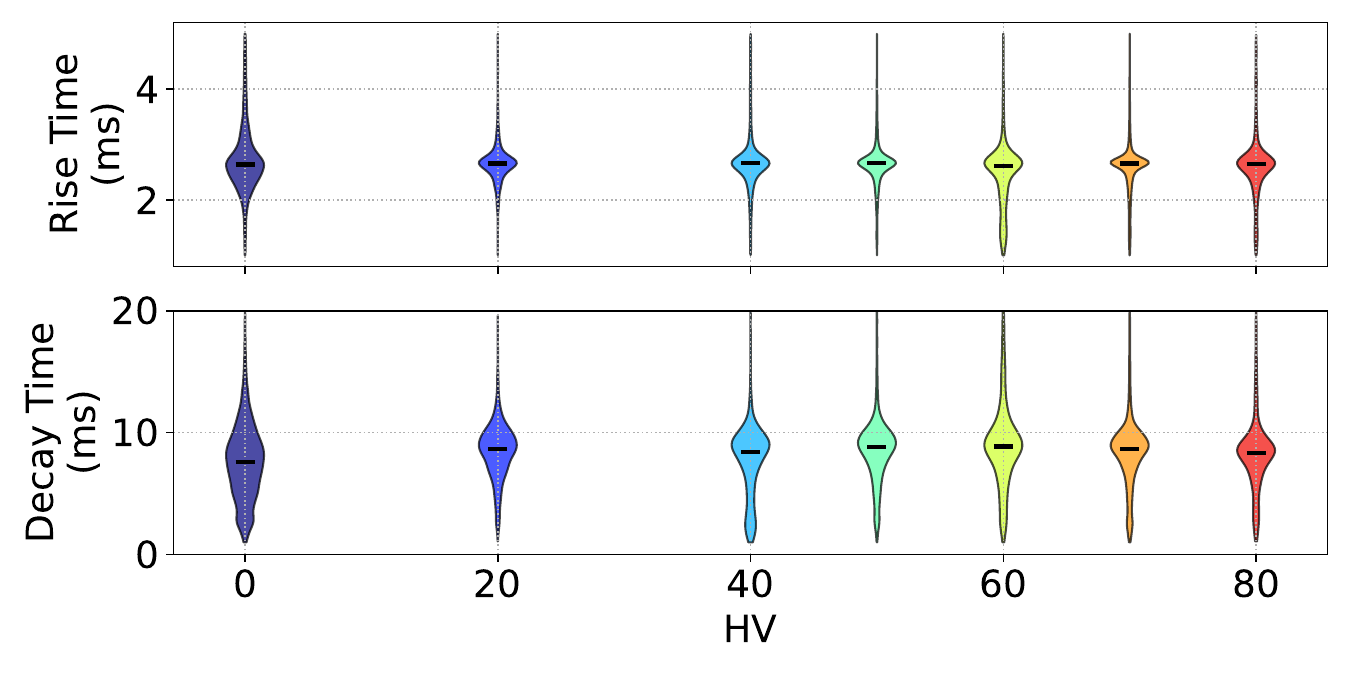}
    \caption{ITO4 rise time and decay time distributions as a function of the applied NTL bias (\VNTL). The average values remain constant for different \VNTL{}, proving that amplification does not change the response of the detector.}\label{fig:time_constants}
\end{figure}

To fully characterize the evolution of the system, the rise and decay times have also been computed for each sub-run, resulting in the distributions plotted for ITO4 in Figure \ref{fig:time_constants}. These quantities, calculated for each measured pulse, parameterize the response function of the cryogenic calorimeter to energy injections, and are connected to the electro-thermal circuit of the detector  \cite{Adams:2022b}. Interestingly, the pulse shape remains  stable at different \VNTL{}, even above the breakdown voltage. The average values for the response times for ITO4 (ITO1) are $2.6 \pm 0.3$ ms ($2.7 \pm 0.4$ ms) for the rise time and $8.4 \pm 2.1$ ms ($10.2 \pm 4.2$ ms) for the decay time.

A constant response time for different \VNTL{} ensures that the NTL gain directly improves the time resolution of the detectors, inversely proportional to product between Signal-to-Noise Ratio (SNR) and rise time \cite{Chernyak2016,Chernyak2012}. Such feature has great importance for the application of these detectors to the CUPID experiment, since one of its expected background components is the pile-up resulting from two $2\nu\beta\beta$ decays  \cite{CUPID:2025faj}. A rejection of this background to negligible values can only be achieved exploiting the light detector pulses, characterized by response times faster than the crystal ones ( $\sim$1~ms against $\sim$10~ms) \cite{Ahmine:2023xhg,Armatol:2021rna}. In this framework, a NTL amplification that does not affect rise time ensures that the light detector can meet the time resolution target. 

\section{Results}\label{results}
\begin{figure}[h]
    \includegraphics[width=0.5\textwidth]{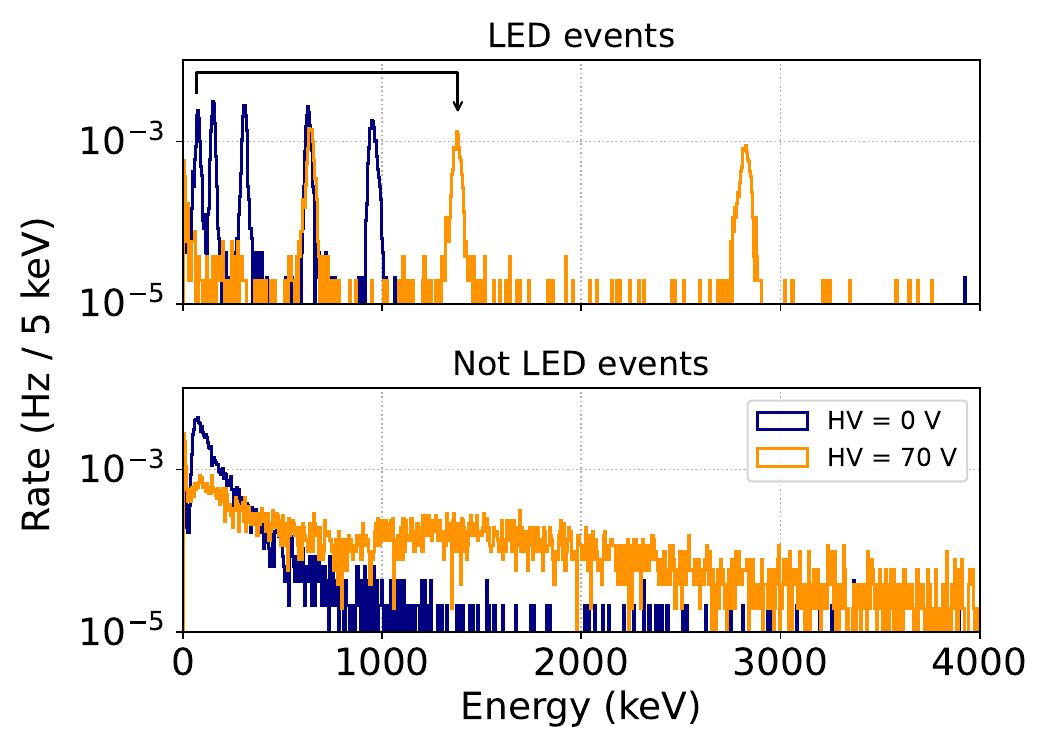}
    \caption{Example of a calibrated spectrum for ITO4. The two event categories, LEDs and (mainly) muons, are separated and plotted for two datasets, \VNTL~=~0~V and \VNTL~=~70~V.} \label{fig:spectrum_mu_LED}
\end{figure}

The main focus of the analysis is the evaluation of the NTL gain.
The method to quantify this parameter relies on measuring the amplitude change after \VNTL{} application for energy depositions of the same intensity.
In the spectra accumulated during the data-taking with the studied detectors, two populations of such events could be identified: muon- and LED- induced events.
The former are expected, and can be tagged requiring time coincidence between at least three detectors (stacked as shown in Figure~\ref{fig:detector}).
The latter are induced in the system and are tagged with a logical flag that marks the time when the LED is fired. In the analyzed measurements, we controlled the LED with a pulse generator, sending square waves with fixed amplitude and 6 different widths: 12~$\mu$s, 25~$\mu$s, 50~$\mu$s, 100~$\mu$s, 200~$\mu$s, 300~$\mu$s. This results in 6 LED pulses with increasing amplitudes.
In Figure~\ref{fig:spectrum_mu_LED}, an example spectrum of ITO4 with the two event categories, with and without NTL voltage, is reported.

\begin{figure}[h]
    \includegraphics[width=0.47\textwidth]{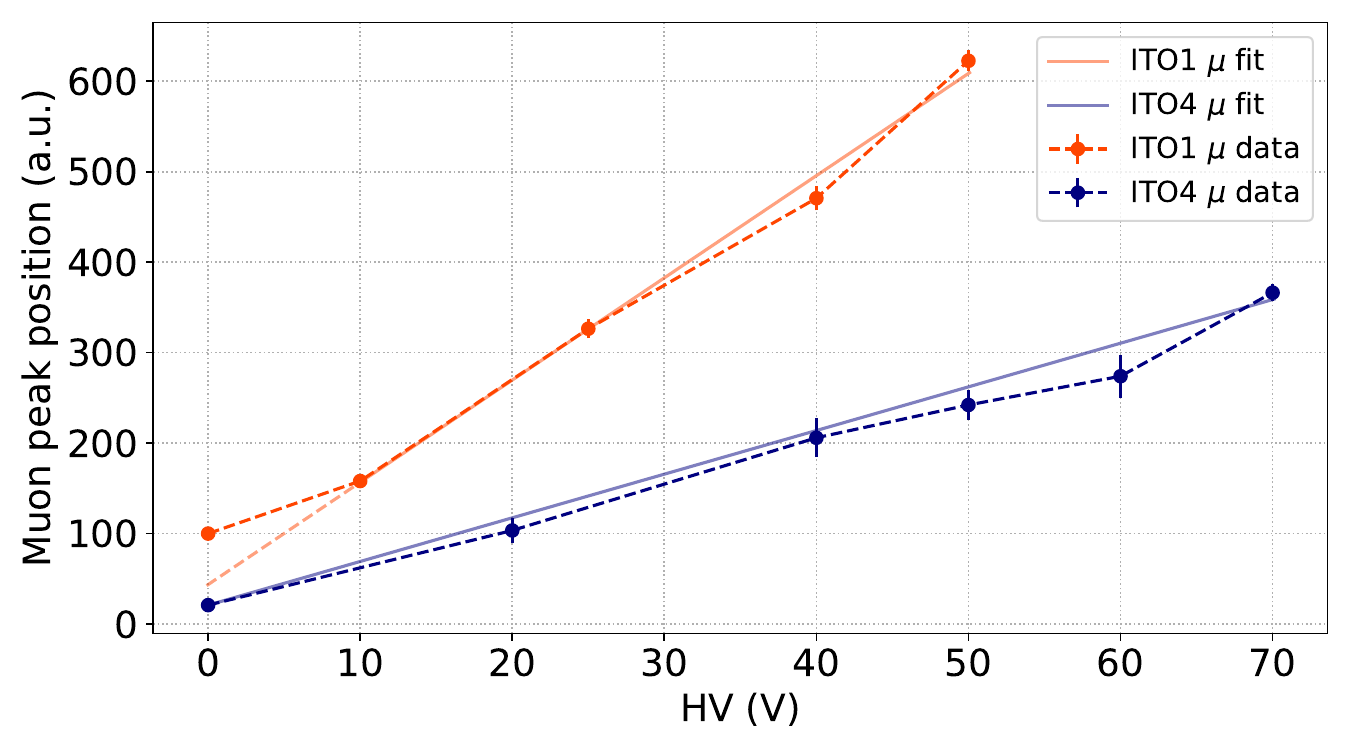}
    \caption{Measured muon peak position for the two working ITO detectors. A linear fit with the expected linear trend is also shown. Both detectors are well represented by the model. The only deviation is for ITO1 with \VNTL{}~=~0~V. In that measurement the muon-induced peak could not be identified, causing the mismatch. By enforcing the model, it is however possible to reconstruct that value extrapolating the gain trend.}\label{fig:mu_gain}
\end{figure}

The upper plot shows the tagged LED events. In the \VNTL~=~0~V dataset, only the five highest peaks are detected, whereas all six peaks are well above the threshold when NTL amplification is applied. A black arrow shows the correspondence between the same peak in the two scenarios. The bottom plot shows non-LED events, which are mainly due to muon interactions. In the HV = 0 V dataset, the event spectrum exhibits a decreasing intensity with increasing energy, primarily due to muon interactions in the silicon wafer. At HV = 70 V, this distribution splits into two populations. One follows a trend similar to the 0 V data, corresponding to unamplified muon events in regions without ITO electrode coverage. The other population develops at energies above 1 MeV and corresponds to amplified events, where muons deposit energy in the ITO-covered areas, generating an NTL-amplified thermal signal.  \\ 
The starting point of the muon population at \VNTL~=~0~V, named $A_\mu^0$, was evaluated by fitting the muon-only spectrum with a Landau distribution.
This quantity corresponds to the minimal expected energy deposition from muons in the detector bulk. Since atmospheric muons are at the ionization minimum ($\tfrac{dE}{d\rho x} \sim 1.66\,\mathrm{MeV\,cm^2/g}$), the minimal energy comes from the shortest muon trajectory in the absorber volume. In our specific setup, this corresponds to a muon perpendicular to the wafer surface, depositing $\sim 78\, \mathrm{keV}$ over \Thickness{} of wafer thickness. For the spectra obtained with \VNTL~\textgreater~0~V, the starting point of the muon population, $A_\mu$, has been instead extracted by fitting with a Landau distribution the population at higher energies, coming from amplified events.  \\
The evolution of $A_\mu$ as a function of \VNTL{} is shown in Figure~\ref{fig:mu_gain}.
This trend is interpolated with the following linear function:
\begin{equation}
    A_\mu = A_\mu^0 \cdot \left(1 + \eta \cdot \frac{e\cdot\VNTL{}}{\omega}\right)
    \label{eq:mu_gain}
\end{equation}
where $A_\mu^0$ is the muon peak position with \VNTL~=~0~V, $\eta$ is the NTL efficiency, $e$ is the elementary charge and $\omega$ is the average energy required to produce an electron-hole pair in the silicon absorber. 
This relation is derived assuming that the energy of the electrical field, $e\cdot\VNTL{}$, is transferred to the thermal channel with an efficiency $\eta$ \cite{Luke:1988}.

By assuming this model, the parameters $A_\mu^0$ and $\eta$ can be estimated from the fit.

In particular, $A_\mu^0$ can be extracted from this interpolation even if the muon peak at \VNTL~=~0~V cannot be identified in the spectra, as in the case of ITO1. Since $A_\mu^0$ is necessary for the energy calibration of a detector, this interpolation allows to recover this information for all devices showing NTL gain. 

The NTL efficiency $\eta$ was estimated as 0.98 (0.88) for ITO 1 (ITO 4) \footnote{These values can be compared with those reported in \cite{Novati:2019}, where values of $\eta$ in the range [0.28, 0.63] were measured. However, it should be noted that the model used to analyze the data is different from the one used in the present work.}.
Values close to 1, as in this case, indicate a low trapping probability during drift, as expected with our electrode geometry, where electrons and holes have to travel for a bulk thickness of up to \Thickness{}.

From the best fit value, it is possible to compute the gain factor for the two detectors, represented by the round bracketed factor of Equation~\ref{eq:mu_gain}. For ITO1 (ITO4), the maximum gain for muons is 14.3 (17.6), corresponding to \VNTL{}~=~50~V (\VNTL{}~=~70~V).     

\begin{figure}
\includegraphics[width=0.53\textwidth]{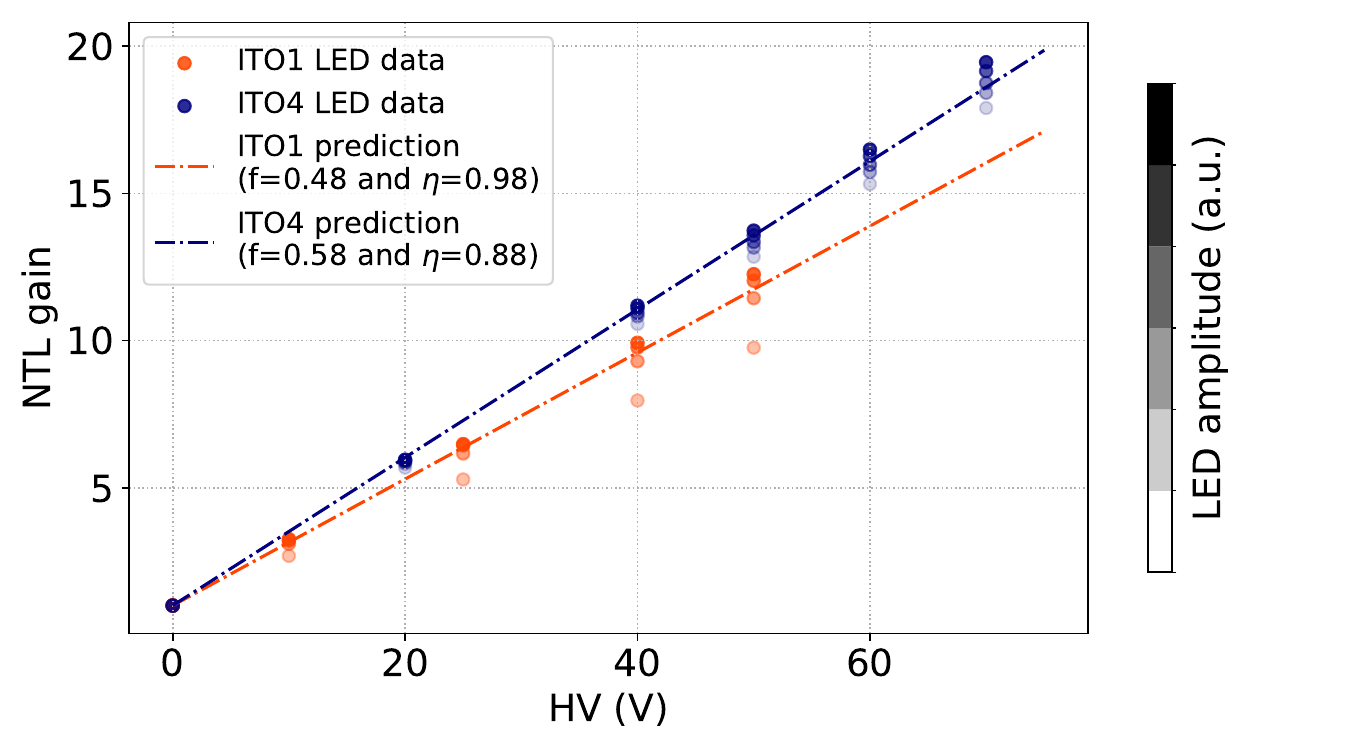}
\caption{Measured gain for LED peaks. A prediction based on an analytical model is also shown. This model explains the difference between the two detector with a different effective coverage of the electrodes, represented by the $f$ factor. The NTL efficiency $\eta$ is fixed to the value extracted by the fit of the gain for muon-induced events.}\label{fig:LED_gain}
\end{figure}

For LED-induced events, the average response to the 6 LED amplitudes for each \VNTL{} was estimated fitting the amplitude spectra with 6 independent Gaussian distributions. The NTL gain as a function of \VNTL{} was then evaluated dividing the averages of the distributions for \VNTL{}~$>$~0~V by those at \VNTL{}~=~0~V.

To model dependence of the gain on \VNTL{} for these events, it is important to note that only the fraction of photons $f$ that interact within the area covered by the electrodes generate an electron-hole pair that is accelerated and contributes to the increase in the thermal signal.
Starting from this consideration, it is possible to derive the following gain function for the LED events, linear in \VNTL{}:
\begin{equation}
    g_{LED} = \frac{f (1 - r_{ITO}) \left(1 + \eta \frac{e \cdot HV}{E_{ph}}\right) + (1 - f) (1- r_{Si}) }{f (1 - r_{ITO}) + (1 - f) (1- r_{Si})}
    \label{eq:LED_gain}
\end{equation}
In this equation, $r_{ITO} = 25\%\, (0\%)$ is the ITO1 (ITO4) reflectivity, which can be obtained from the data shown in Figure \ref{fig:reflectivity}, $r_{Si} = 37\%$ is the silicon reflectivity, $E_{ph} = 2.4\, eV$ is the optical photon energy and $\eta$ is the NTL efficiency, as in Equation~\ref{eq:mu_gain}.
Assuming full coverage of the detector area ($f=1$), Equation~\ref{eq:LED_gain} reverts to that usually found in the literature:
\begin{equation}
    g_{LED} = 1 + \eta \cdot \frac{e \cdot HV}{E_{ph}}
\end{equation}
which assumes that every absorbed optical photon produces an electron-hole pair that is drifted along the electrical field with efficiency $\eta$.
The measured NTL gains for the 6 LED amplitudes are shown in Figure~\ref{fig:LED_gain} for each \VNTL{} value. The dashed line in the same plot is the  theoretical prediction for the trend, based on Equation~\ref{eq:LED_gain}.
The value for $\eta$ is extracted from the fit of Equation~\ref{eq:mu_gain} to muon data (Figure~\ref{fig:mu_gain}). This assumption is based on the fact that the path for drifting electron-hole pairs is the same in both cases.
The value for $f$ instead is obtained from geometrical considerations. Assuming a uniform illumination of the light detector, $f$ is expected to be the ratio between electrode area and Si surface, which is 0.58.
This value yields a theoretical prediction that is consistent with the measured data for ITO4 that, being the highest in the stack of devices operated in the cooldown (see Figure~\ref{fig:detector}) is compatible with the hypothesis of uniform illumination. For this detector, a maximum NTL gain of 19.46 was achieved, corresponding to \VNTL{}~=~70~V.
On the other hand, reproducing the ITO1 data requires a value of $f \sim 0.48$. This effective reduction in coverage is justified by the fact that ITO1 is at the bottom of the detector stack. Consequently, uniform illumination of the electrode surface is less likely and the interaction with photons on the outer, uncovered ring is more probable. For this device, a maximum gain of 12.25 was achieved, corresponding to \VNTL{}~=~50~V. The extracted gain for LED events is consistent with that measured for the same voltages using CUPID-like NTL light detectors \cite{Novati:2019}. 

We also observed an unexpected $\sim 10 \%$ dependence of the NTL gain on the LED amplitude, which is directly proportional to the number of photons interacting with the detector.
The physical reason for this dependency is still under investigation and will be the focus of future studies.

The independent evaluation of the gain on muon- and LED-induced event is summarized in Table~\ref{tab:NTL_gains}. The two values are extracted from the best fits reported in Figure~\ref{fig:mu_gain} and \ref{fig:LED_gain} and are not compatible with each other.

\begin{table}
% table caption is above the table
\caption{Maximal NTL gain values for the tested detectors}\label{tab:NTL_gains}%
% For LaTeX tables use
\begin{tabular}{cccc}%{0.5\textwidth}{@{\extracolsep{\fill}}cccc@{}}
\hline\noalign{\smallskip}
Detector    &   Max \VNTL{} [V] &   Gain Muons  &   Gain LED    \\
\noalign{\smallskip}\hline\noalign{\smallskip}
ITO1    &   50  &   14.3$\pm$0.3  &   12.25$\pm$0.01  \\
ITO4    &   70  &   17.6$\pm$0.5  &   19.46$\pm$0.01  \\
\noalign{\smallskip}\hline
\end{tabular}
\end{table}

\begin{figure}[h]
    \includegraphics[width=0.47\textwidth]{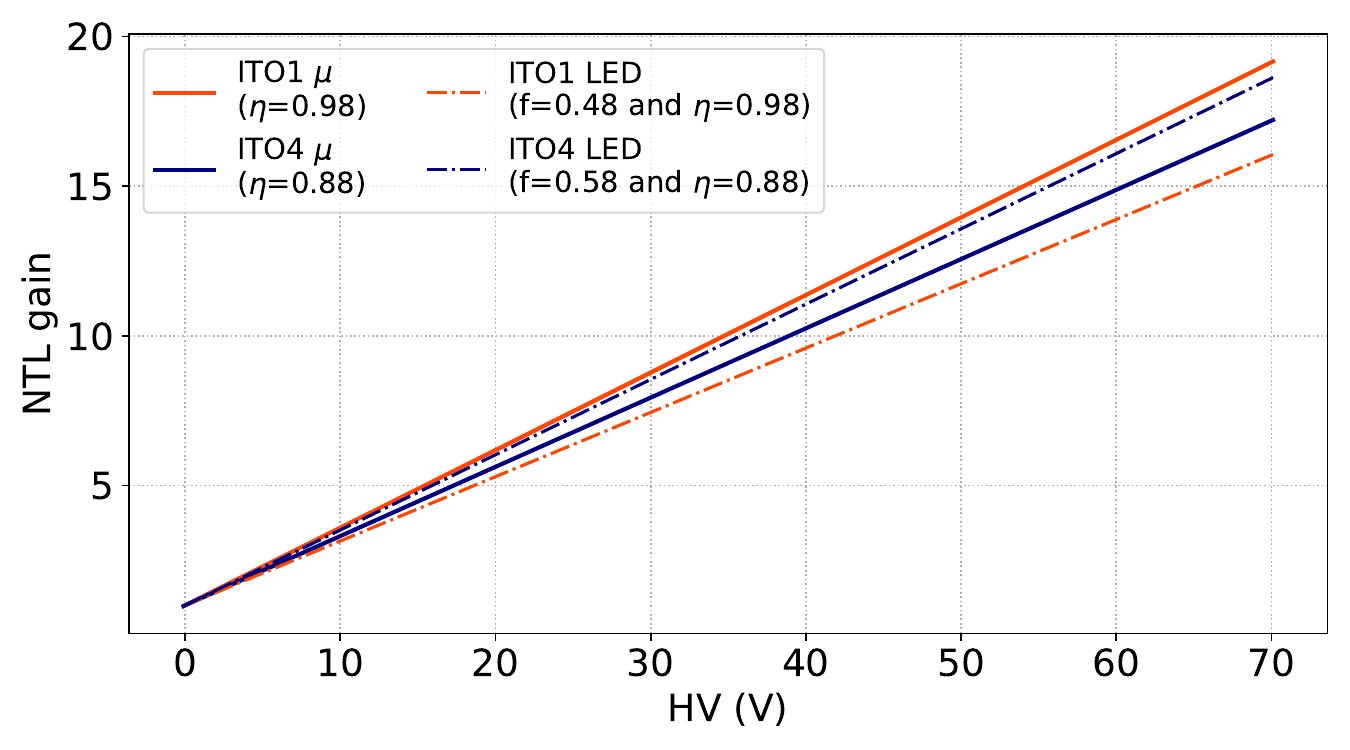}
    \caption{Comparison of the observed NTL gain functions for muon- (solid lines) and LED-induced (dash-dotted lines) events. The two curves are different for both the detectors, but are modeled by the two proposed models. This double modeling is fundamental to export calibration from ionization- to light-induced events.}\label{fig:all_gain}
\end{figure}

To better elaborate on this difference, a summary plot comparing the gain functions measured for LED- and muon-induced events as a function of \VNTL{} for the two detectors characterized in this work is reported in  Figure \ref{fig:all_gain}. Both series of values are computed from the best-fit function in Figure \ref{fig:mu_gain} and \ref{fig:LED_gain}.
The two detectors present differences in gain that the model we used traces to a different drift efficiency ($\eta$) and a different spatial distribution of the incoming photons.
For both detectors, however, LED- and muon-induced events present different gain curves, well described independently by different models (Equations \ref{eq:mu_gain} and \ref{eq:LED_gain}). 

This latter difference has crucial implications for the operation and design of these devices. 
One of the main advantages of an NTL-assisted light detector with a gain for ionizing particle events is the ability to calibrate the device in energy, even with NTL amplification.
However, since the ionization and light gain curves differ, this energy calibration can only be translated to light events without bias if the spatial distribution of photons, represented by the $f$ factor of Equation~\ref{eq:LED_gain}, is known precisely. 
Only under this assumption can the two different gain vs \VNTL{} curves be used to transfer the calibration from \VNTL{}~=~0~V to the operational \VNTL{}.
Accurately and precisely evaluating $f$ can be challenging when working with scintillation light from crystals, due to the many parameters affecting the shape of the scintillation photon distribution.
Increasing the NTL electrode coverage could mitigate this issue. Future ITO light detector design and characterization work will focus on this task.

\section{Conclusions and outlook}\label{conclusions}
In this work the characterization of two silicon cryogenic light detectors with transparent ITO electrodes for NTL amplification has been presented.
ITO transparent electrodes allow to generate an electric field perpendicular to the silicon wafer.
This is a novel and promising configuration of the electric field compared to traditional implementations where the electric field is parallel to the wafer surface.
The measured NTL gain is very close to the expected theoretical one.
Specifically, we developed a consistent analytical model to explain the measured gain for both minimum ionizing particles (muons) and optical photons that takes into account the partial coverage of the device surface by the electrodes.
A demonstrated gain of $\sim$20 for one of the two tested devices, the easy and inexpensive realization procedure and the fact that ITO electrodes double as anti-reflective coating already make these device a promising solution for next generation projects needing cryogenic light detectors with close to single-photon sensitivity.
Future productions aim at testing the repeatability of the production process and long term operation.
Further optimizations include increasing electrodes coverage, test different substrate materials and dependence of maximum bias voltage on wafer thickness.

%\begin{acknowledgements}
%If you'd like to thank anyone, place your comments here
%and remove the percent signs.
%\end{acknowledgements}

% BibTeX users please use one of
%\bibliographystyle{spbasic}      % basic style, author-year citations
%\bibliographystyle{spmpsci}      % mathematics and physical sciences
\bibliographystyle{spphys}       % APS-like style for physics
\bibliography{references}   % name your BibTeX data base

\end{document}